\documentclass[10pt,conference]{IEEEtran}

\usepackage{cite}
\usepackage{amsmath,amssymb,amsfonts}
\usepackage{amsthm}
\usepackage{graphicx}
\usepackage{textcomp}
\usepackage{algorithm}
\usepackage{algorithmic}
\usepackage[caption=false,font=footnotesize]{subfig}
\usepackage[hidelinks]{hyperref}

\renewcommand{\qedsymbol}{$\blacksquare$}

\newtheorem{definition}{Definition}
\newtheorem{coro}{Corollary}
\newtheorem{thm}{Theorem}
\newtheorem{lemma}{Lemma}
\newtheorem{remark}{Remark}

\newcommand{\figref}[1]{Fig.~\ref{#1}}
\newcommand{\algorithmref}[1]{Algorithm~\ref{#1}}
\newcommand{\theoremref}[1]{Theorem~\ref{#1}}
\newcommand{\sectionref}[1]{Section~\ref{#1}}

\def\BibTeX{{\rm B\kern-.05em{\sc i\kern-.025em b}%
\kern-.08emT\kern-.1667em\lower.7ex\hbox{E}\kern-.125emX}}

\title{Fundamental Limits of Man-in-the-Middle Attack Detection in Model-Free Reinforcement Learning}

\author{
Rishi Rani, \IEEEmembership{Student Member, IEEE},
Massimo Franceschetti, \IEEEmembership{Member, IEEE}
\thanks{The simulation code repository can be found at: \url{https://github.com/rishirani/RL_MDP_MITM}.}
\thanks{Rishi Rani is a Ph.D. candidate with the Department of Electrical and Computer Engineering, University of California San Diego, La Jolla, CA 92093 USA (e-mail: smr@ucsd.edu).}
\thanks{Massimo Franceschetti is with the Department of Electrical and Computer Engineering, University of California San Diego, La Jolla, CA 92093 USA (e-mail: massimo@ece.ucsd.edu).}
}

\begin{document}

\maketitle

\begin{abstract}
We consider the problem of learning-based man-in-the-middle (MITM) attacks in cyber-physical systems (CPS), and extend our previously proposed Bellman Deviation Detection (BDD) framework for model-free reinforcement learning (RL). We refine the standard MDP attack model by allowing the reward function to depend on both the current and subsequent states, thereby capturing reward variations induced by errors in the adversary’s transition estimate. We also derive an optimal system-identification strategy for the adversary that minimizes detectable value deviations. Further, we prove that the agent’s asymptotic learning time required to secure the system scales as \(\Theta(t_b)\), where \(t_b\) is the adversary’s learning time, and that this matches the optimal lower bound. Hence, the proposed detection scheme is order-optimal in detection efficiency. Finally, we extend the framework to asynchronous and intermittent attack scenarios, where reliable detection is preserved.
\end{abstract}


\section{Introduction}

\label{sec:introduction}

Recent advancements in wireless technology and computational capabilities have made it possible to perform networked control in cyber-physical systems (CPS), enabling a wide range of applications such as cloud robotics, autonomous navigation, and industrial process management~\cite{Surv}. These systems are inherently online, making real-time decisions based on past observations in a closed-loop fashion. However, the distributed architecture of CPS introduces significant security vulnerabilities, necessitating the development of secure and optimal control strategies. Security breaches in these systems can have catastrophic consequences, including attacks on financial systems, hijacking of autonomous vehicles and unmanned aerial vehicles, or compromising life-critical infrastructure~\cite{Attack1, Attack2, Attack3}. High-profile incidents such as the cyber-attacks on the Ukraine power grid, the German steel mill, the Australian sewage system, the Davis-Besse nuclear power plant, and the Stuxnet malware attack on the Iranian uranium enrichment facility highlight the severity of these threats~\cite{RWexamples}. These events have spurred extensive research into the prevention of security breaches at a control-theoretic level~
\cite{MLBA7, MLBA10, MLBA13}.

Within this context, the “man-in-the-middle” (MITM) attack model is a significant concern in CPS. In such attacks, an adversary intercepts and manipulates sensor feedback signals sent from the physical plant to the legitimate agent, feeding spoofed signals that appear to indicate safe and stable operation. Meanwhile, the adversary also manipulates the control signal to steer the plant toward a catastrophic state. To detect these attacks, the legitimate agent must continuously monitor plant outputs and search for statistical anomalies in the feedback. Conversely, the attacker aims to generate spoofed sensor data that is statistically indistinguishable from the legitimate signals while driving the system towards failure.

Two prominent MITM attack types have been studied in detail. The first, the \textit{replay attack}, involves the adversary recording system behavior over time and then replaying it periodically to deceive the agent~\cite{MLBA26, MLBA27}. This attack is easier to detect, as it does not require knowledge of system parameters. One countermeasure is to embed a watermark signal into the control input, making it undetectable to the adversary~\cite{MLBA30, MLBA34}. The second, the \textit{statistical-duplicate attack}, assumes the adversary has complete knowledge of the system’s dynamics and parameters, allowing them to construct trajectories statistically identical to the true system’s behavior~\cite{SMITH201190, MLBA29, MLBA30}. Detecting this attack is more complex due to the adversary’s full system knowledge, requiring sophisticated detection techniques. To counter this, agents can employ methods such as \textit{moving target}\cite{MLBA35, MLBA36, MLBA37, MLBA38}, \textit{baiting}\cite{MLBA39, MLBA40}, or private randomness through watermarking~\cite{MLBA29}.

Another category of MITM attacks, \textit{learning-based attacks}, arises from the broader field of learning-based control~\cite{LBA1,LBA2,LBA5}. In this case, the adversary lacks initial knowledge of the system but can learn its dynamics over time through observation. This type of attack is more practical, as perfect knowledge of the system is unrealistic. However, once the adversary learns the system model, they can launch sophisticated deception schemes. Previous studies have used information-theoretic approaches to derive bounds on the probability of successful deception in scalar and vector linear time-invariant systems~\cite{Mohammad_LBA}, as well as bounds on the agent’s time and energy required to detect the attack with confidence~\cite{Rangi_LBA}.

In our previous work\cite{MDP_conf, LQR_conf}, we extend the model of learning-based attacks to include the learning process of the agent itself. Specifically, we focus on a legitimate agent that performs model-free control through reinforcement learning (RL). In this setting, where the agent has no explicit system model, attack detection becomes particularly challenging. We propose a novel attack detection algorithm, the ``Bellman Deviation Detection” (BDD) algorithm, which guarantees detection with high probability while avoiding false alarms, provided an “information advantage” condition is met. This condition links the error in the agent’s Q-function to the adversary’s error in modeling the system. Furthermore, our analysis provides insights into the information patterns required for successful detection. 

\subsection{Contributions}

This paper extends our prior work on the Bellman Deviation Detection (BDD) algorithm for Markov decision processes~\cite{MDP_conf}. Our main novel contributions are as follows:
\begin{enumerate}
    \item \textbf{Refined Problem Formulation.} We refine the standard MDP attack model by allowing the reward function to depend not only on the current state and action but also on the subsequent state. This captures the fact that, under a man-in-the-middle (MITM) attack, the reward statistics can drift as a function of the underlying state transition model. Building on this more nuanced formulation, we derive a modified version of the original BDD algorithm that explicitly accounts for changes in the transition dynamics of the underlying system.
    
    \item \textbf{Adversary’s Optimal System Identification.} We devise a simple system-identification (SI) algorithm and prove that it is optimal in terms of minimizing the deviation in value estimates that the agent uses to detect an attack. The resulting SI algorithm yields the system estimate that produces the smallest detectable trajectory deviations.
    
    \item \textbf{Extension to Asynchronous and Intermittent Attacks.} The original BDD framework assumed a synchronous, phased attack schedule. Here, we generalize to a \emph{dynamic} attack model in which the adversary may attack \emph{asynchronously} and on an \emph{intermittent} basis. We show how to adapt the BDD algorithm to this setting and prove that it still reliably detects attacks even when they occur sporadically.
    
    \item \textbf{Detection Complexity and Cost–Tradeoff Analysis.} We compare the sample complexity—i.e., the learning time required by both the adversary (to mount an undetectable attack) and the agent (to detect it). By deriving a notion of \emph{asymptotic detection complexity}, we prove that the modified BDD algorithm remains order-optimal in the agent’s learning time and matches the order of detection efficiency of model-based detectors.
    
%    \item \textbf{Safety Dual and Model-Free Guarantees.} Finally, we observe that a system whose model parameters drift after deployment can be seen as the \emph{safety dual} of the MITM attack problem. Leveraging this insight, we devise a “Bellman Deviation Estimation” (BDE) algorithm that provides model-free safety guarantees for reinforcement learning. If the underlying system drifts after training, the BDE algorithm detects unsafe drift in real time.
\end{enumerate}

\section{Mathematical Preliminaries and Notation}
A Markov Decision Process is defined by the quadruple $(\mathcal{X}, \mathcal{U}, \mathbf{P}, \mathbf{R}, \gamma)$, where $\mathcal{X}$ is the set of states with cardinality $|\mathcal{X}| = N$ and $\mathcal{U}$ is the set of actions with cardinality  $|\mathcal{U}| = M$. $\mathbf{P}$ is the transition probability matrix, $\mathbf{R}$ is the reward matrix and $\gamma$ is the discount factor. The probabilistic transitions from state to state are Markov and are given by
\begin{gather}
    \Pr(x_{t+1} \mid x_t, u_t) = \mathbf{p}_{x_t,u_t} \equiv [p_{x_t,u_t}(x_1), \dots, p_{x_t,u_t}(x_N)], \\ \nonumber
    \text{and } \mathbf{P} = 
    \begin{bmatrix}
    \mathbf{p}_{x_1,u_1} \\
    \vdots \\
    \mathbf{p}_{x_N,u_M}
    \end{bmatrix},
\end{gather}
where the rows are indexed over all $(x, u) \in \mathcal{X} \times \mathcal{U}$. Similarly, the reward for each transition is given by
\begin{gather}
    r(x_t,u_t, x_{t+1}) = \mathbf{r}_{x_t,u_t} \equiv [r_{x_t,u_t}(x_1), \dots, r_{x_t,u_t}(x_N)], \\ \nonumber
    \text{and } \mathbf{R} = 
    \begin{bmatrix}
    \mathbf{r}_{x_1,u_1} \\
    \vdots \\
    \mathbf{r}_{x_N,u_M}
    \end{bmatrix}.
\end{gather}
The model-free control objective is to learn a policy function $\pi(x) : \mathcal{X} \to \mathcal{U}$ such that the following discounted reward is maximized:
\begin{equation}
    \pi^*(x) = \text{arg}\max_{\pi} \mathbb{E}\left[ \sum_{t=0}^\infty \gamma^t r(x_t, \pi(x_t), x_{t+1}) \right], \quad x_0 \in \mathcal{X},
\end{equation}
where the discount factor, $\gamma$, represents how much the future reward is discounted. This problem is termed the \textit{infinite time horizon discounted reward problem.} This objective is achieved by learning the optimal Q-function of the problem, which is
\begin{gather}
    Q^*(x,u) = \max_{\pi} \mathbb{E} \left[ r(x_0,u_0, x_1) + \sum_{t=1}^\infty \gamma^t r(x_t, \pi(x_t), x_{t+1}) \right], \\ \nonumber
    \text{where } x_0 = x \text{ and } u_0 = u.
\end{gather}
The optimal Q-function relates to the optimal policy as $\pi^*(x) = \text{arg}\max_u Q^*(x,u)$, and the optimal value function, which describes the total accrued reward of an optimal trajectory, is defined as
\begin{gather}
    V^*(x) = \max_u Q^*(x,u), \\ \nonumber
    \mathbf{v} = [V^*(x_1), \dots, V^*(x_N)],
\end{gather}
where $\mathbf{v}$ denotes the optimal value function as a vector. Finally, we note that the optimal Q-function can be recursively written using the \textit{Bellman equation} as
\begin{align}
\label{eq:BellmanEq}
    Q^*(x,u) = & \sum_{x' \in \mathcal{X}} p(x,u,x') \left( r(x,u,x') + \gamma \max_{u'} Q^*(x',u') \right) \\ \nonumber
    = & \sum_{x' \in \mathcal{X}} p(x,u,x') \left( r(x,u,x') + \gamma V^*(x') \right) \\ \nonumber
    = & \mathbf{p}_{x,u} \left( \mathbf{r}_{x,u}^T + \gamma \mathbf{v}^T \right).
\end{align}
Throughout the paper, we describe vectors using boldface, and vectors are defined as row vectors by default (to align with MDP conventions). Matrices are boldfaced and capitalized, and $\|\cdot\|_2$ refers to the Euclidean norm. Finally, we say that an event occurs with high probability (w.h.p.) if its probability \( p_n \) tends to 1 as \( n \to \infty \). Note that proofs of all theorems presented are deferred to the appendix.
\section{Problem Statement}
\label{sec:PS}

The system is modeled as a Markov Decision Process (MDP) controlled by an agent receiving a reward corrupted by additive noise. The reward noise $w_t$ is assumed to be i.i.d. with zero mean; notably, we do not assume finite variance, allowing for heavy-tailed distributions. While standard RL analyses often rely on finite-variance assumptions to invoke the Central Limit Theorem, our framework is designed to be robust even when the variance is infinite. This represents an analytical strength rather than a limitation, as it ensures that the Bellman Deviation Detection (BDD) remains valid for a broader class of noise processes, including those with impulsive characteristics common in networked control systems. We assume that the agent has learned an estimate of the optimal Q-function $\hat{Q}$ using a training trajectory $\boldsymbol{\tau}_a$ described as

\begin{equation}
    \boldsymbol{\tau}_{A}= (x_1, u_1, \dots x_{t_a}, u_{t_a}),
\end{equation}
where $t_a$ is the agent training time. No additional assumption is made on $\boldsymbol{\tau}_a$ itself and the trajectory can be controlled by the agent. The agent has no information about the system model or reward function and uses a generalised learning algorithm with the following stochastic guarantee,

    \begin{gather}
     \label{eq:ConvRate}
    |\hat Q_{t_a}(x,u)- Q^* (x,u)|\leq \epsilon(t_a), \text{w.h.p}\\ \nonumber \text{and } \forall x \in \mathcal{X} , u \in \mathcal{U} \\ \nonumber
    \text{ s.t } \epsilon (t_a)\rightarrow 0 \text{ as } t_a \rightarrow \infty.
\end{gather}

\begin{figure} 
    \centering
  \subfloat[a\label{fig:learning phase}]{%
       \includegraphics[width=0.95\linewidth]{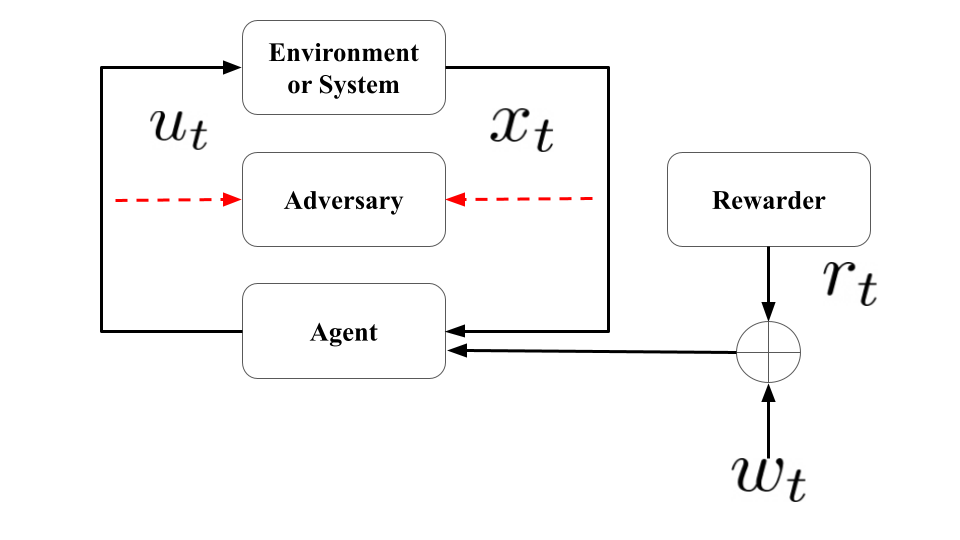}}
    \hfill
  \subfloat[b\label{fig:attack phase}]{%
        \includegraphics[width=0.95\linewidth]{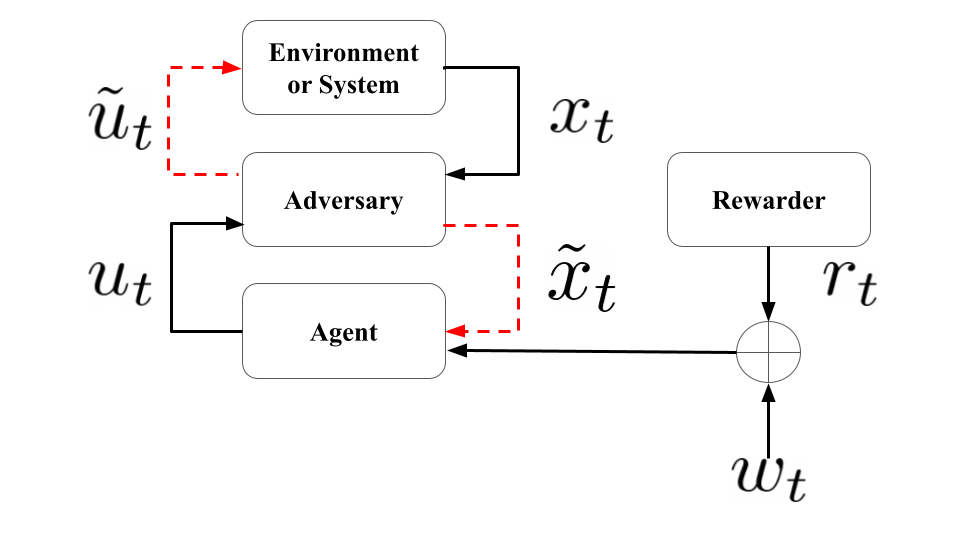}}
    \\
 
 \caption{(a) \textbf{Adversary Learning Phase:} During this phase, the attacker eavesdrops and learns the system dynamics without altering the feedback signal to the agent. (b) \textbf{Adversary Attack Phase:} During this phase, the attacker intercepts the feedback loop and provides a falsified signal to the agent to induce a target policy or cause value deviation.}
  \label{fig1} 
\end{figure}

As described in \figref{fig:learning phase}, the adversary initially is in its learning phase where it observes a trajectory $\boldsymbol{\tau}_b$ and it learns the system giving it an estimate of the transition model $\mathbf{ \hat P}$.  During its learning phase  the adversary has no control over its learning trajectory $\boldsymbol{\tau}_b$, as it merely learns by observing and does not control the system. Therefore, no asymptotic convergence guarantees are placed on its estimate $\mathbf{\hat P} $. In the attack phase (as described in \figref{fig:attack phase}) the agent takes control of the system and feeds the agent a spoofed state feedback signal. This feedback signal is statistically consistent with its transition model estimate $\mathbf{\hat P}$. Note that $\mathbf{\hat P}$ need not be an explicit estimate made by the adversary (for example the adversary may also use model-free learning), however there exists an implicit statistical model it follows. The trajectory $\boldsymbol{\tau}_c$ formed during the attack phase  is used by the agent to perform attack detection. The adversary in this phase steers the true system towards catastrophe and the agent is tasked with detecting the attack and declaring a breach.  The   adversary's strategy to lead the system to catastrophe does not affect attack detection, namely the adversary's closed feedback with system is not of strict concern to the detection problem.
\newline \newline
\textbf{Problem Statement:} Given the agent has a learned estimate of the optimal Q-function $\hat Q(\cdot)$ from the trajectory $\boldsymbol{\tau}_a$ and the adversary spoofs the system with a transition model estimate $\mathbf{\hat P}$, devise a detection algorithm that uses the trajectory during attack $\boldsymbol{\tau}_c$ and provides guarantees on attack detection as the trajectory length $t_c \rightarrow\infty$. 

\section{A Detection Algorithm Based on ``Bellman Deviation''}
In this section, we describe our proposed algorithm and prove its stochastic guarantees.

\subsection{Algorithmic Description}

Before we describe the detection algorithm, we begin by defining the key quantities required for our analysis. The trajectory observed during an attack is a tuple of the form 
\[
\boldsymbol{\tau}_c = (x_1, u_1, \dots, x_{t_c}, u_{t_c}).
\]
Let \( t_c(i,j) \) be the number of times the state-action pair \( (i,j) \) is observed, and the sequences \( x_{i,j}(k) \) and \( u_{i,j}(k) \) represent the respective states and actions that followed them each subsequent time. Similarly, let \( r_{i,j} \) be the immediate reward doled out at that instant, and \( w_{i,j}(k) \) be its associated white noise.

\begin{definition}[Bellman Deviation Process]
Let 
\begin{gather}
d_{i,j}(k) = \hat{Q}(i,j) - r_{i,j} - w_{i,j}(k) - \gamma \hat{V}(x_{i,j}(k)) \\ \nonumber \forall k \in [1,t_c(i,j)], 
\end{gather}
be the Bellman deviation process. This sequence represents the deviations from Bellman-consistent behavior in the observed trajectory during the attack phase.
\end{definition}

The Bellman deviation process (BDP) is simply the empirical temporal difference (TD) error, separated by state-action pair, to form \( M \times N \) different sequences, each representing the sequence of TD errors measured in the trajectory when the system transitioned through the respective state-action pair.

\begin{definition}[Bellman Deviation Process Mean]
We define
\begin{align}
\label{eq:deviation average}
\bar{d}_{i,j} = \frac{\sum_{k=1}^{t_c(i,j)} d_{i,j}(k)}{t_c(i,j)}
\end{align}
to be the Bellman deviation process mean (BDPM). The (BDPM) is simply the sample mean of the BDP. 
\end{definition}

Taking the sample mean reduces the effect of disturbances due to noise in rewards and stochastic transitions, and is an effective metric to detect deviations from Bellman-consistency. We use bounds on the BDPM to determine if the system is under a MITM attack. A high BDPM would suggest that the system is under attack. However, to establish precise bounds on the deviation averages, we need to define an informative metric on the adversary's model error to characterize its detectability.

\begin{definition}[Minimum Value Drift]
Given an MDP system \( (\mathcal{X}, \mathcal{U}, \mathbf{P}, \mathbf{R}) \) and the adversary's system model \( \mathbf{\hat{P}} \), we can define its minimum value drift as
\begin{align}
\Phi_\text{min} = \min_{i,j \in \mathcal{X}\times \mathcal{U}} |\mathbf{\tilde p}_{i,j} (\mathbf{r}_{i,j} + \gamma \mathbf{v}) |
\end{align}
where \( \mathbf{v} \) is a matrix with the optimal value row vector, and \( \mathbf{\tilde{p}}_{i,j} \) is the $(i,j)$ indexed row from the adversary's model error \( \mathbf{\tilde{P}}  = \mathbf{{P}} - \mathbf{\hat{P}} \).
\end{definition}

The above definition can be understood intuitively as a measure that tells us how easily we can observe statistical deviations in the system's trajectory during the attack phase. For example, if the system has \( \Phi_\text{min} = 0 \), this implies that the value function gives us no information about the different trajectories, as there is no detectable minimum drift. Therefore, the minimum value drift measure is a key feature of the system and should be evaluated when designing secure systems.  

With the above quantities defined, we are now ready to present the Bellman deviation detection algorithm (see Algorithm \ref{alg:BellDev}) and prove its correctness.

\begin{algorithm}
\caption{Bellman Deviation Detection}\label{alg:BellDev}
\begin{algorithmic}
\REQUIRE \( t_c \geq 0 \), length(\( \mathbf{\tau}_C \)) = \( t_c \), \( \gamma \)
\STATE \( \bar{\epsilon} \geq \max_{i,j} |Q(i,j) - \hat{Q}(i,j) | \) w.h.p. 
\STATE \textbf{initialize} \( \mathbf{D} \gets \mathbf{[0]} \)
\STATE \textbf{initialize} \( \mathbf{T} \gets \mathbf{[0]} \)
\STATE \textbf{initialize} \( n \gets 1 \)

\WHILE{\( n \leq t_c \)}
\STATE \( i \gets \mathbf{\tau}_C[n][0] \)
\STATE \( j \gets \mathbf{\tau}_C[n][1] \)
\STATE \( k \gets \mathbf{\tau}_C[n+1][0] \)
\STATE //\textit{Accumulating the values of the BDP.}
\STATE \( \mathbf{D}[i,j] \gets \hat{Q}(i,j) - r(i,j,k) - \gamma \hat{V}(k) + \mathbf{D}[i,j] \)
\STATE \( \mathbf{T}[i,j] \gets \mathbf{T}[i,j] +1 \)
\STATE \( n \gets n + 1 \)
\ENDWHILE
\STATE //\textit{Pruning the BDPs whose means have not converged}
\STATE PRUNE$(\mathbf{D},\mathbf{T})$
\STATE //\textit{Calculating BDPM by dividing by times visited}
\STATE \( \mathbf{D} \gets \frac{\mathbf{D}}{\mathbf{T}} \)
\STATE //\textit{Threshold largest BDPM by security bound}
\IF{ \( \max(\mathbf{|D|}) > (1 + \gamma)\bar{\epsilon} \)} 
\STATE \text{declare breach}
\ELSE
\STATE \text{declare no breach}
\ENDIF
\newline
\STATE //\textit{This function prunes BDPs with unconverged means, i.e., infrequently visited state-action pairs}
\STATE \textbf{function} PRUNE$(\mathbf{D},\mathbf{T})$
\STATE \textbf{initialize} $m = \frac{1}{2\cdot|\mathcal{X}||\mathcal{U}|}$
\FORALL{$(i,j)$ \textbf{in} $\mathcal{X}\times \mathcal{U}$}
    \IF {$\mathbf{T}[i,j] \leq m\cdot t_c$}
        \STATE DELETE$(\mathbf{D}[i,j],\mathbf{T}[i,j])$ 
    \ENDIF
\ENDFOR
\STATE \textbf{end function}
\end{algorithmic}
\end{algorithm}

In Algorithm \ref{alg:BellDev}, the division \( \frac{\mathbf{D}}{\mathbf{T}} \) is an element-wise division of the two matrices. The algorithm essentially calculates the BDPMs \( \bar{d}_{i,j} \), and prunes out the BDPMs that were not sufficiently averaged and have therefore not converged yet. Then it takes the maximum absolute value among them and compares it with the security bound \( \xi = (1 + \gamma) \bar \epsilon \). If it exceeds this bound, a breach is declared. Note that the algorithm guarantees attack detection and no false alarms, with high probability, if and only if the information advantage condition is met. The information advantage condition essentially compares the agent's information about the system to the adversary's by comparing their learning errors, and the above algorithm is only effective if the agent has greater information about its environment than the adversary to avoid deception. It is fully proven and described in the following subsection.

\begin{remark}
We point out that the algorithm does not require exact estimates of the error bound of the Q-function \( \epsilon(t_a) \), but rather an overestimate (\( \bar\epsilon \)). This allows for more practical scenarios where an exact value of the quantity would be unavailable and could be obtained through bootstrap methods with certain confidence bounds.
\end{remark}

\subsection{Correctness of the Algorithm}

In this section, we prove the correctness of the proposed algorithm. We begin by proving an asymptotic upper bound on the BDPMs when no attack is underway. To this end, we derive the asymptotic limit of the BDPMs under secure conditions.

\begin{thm}
\label{Thm: lim1}
In the absence of attacks, the BDPMs converge asymptotically as,
\begin{gather}
\label{eq: proof2}
    \bar{d}_{i,j} \rightarrow  \hat{Q}(i,j) - Q^*(i,j) + \gamma  \mathbf{p}_{i,j} \left(   \mathbf{v} - \mathbf{\hat{v}} \right)^T, \\ \nonumber
    \text{w.h.p. as } t_c(i,j) \rightarrow\infty, 
    \; \forall (i,j) \in \mathcal{X} \times \mathcal{U}.
\end{gather}
\end{thm}
We now use the asymptotic limit to prove an asymptotic upper bound on the magnitude of the BDPM when no attack is underway.

\begin{thm}
\label{thm:UB}
In the case when no attack occurs, the magnitude of the BDPMs can be upper-bounded as,
\begin{gather}
\label{eq: proof2}
    | \bar{d}_{i,j} | \leq (1+\gamma) \epsilon(t_a), \text{w.h.p. as } \\ \nonumber
    t_c(i,j) \rightarrow \infty, \;\; \forall (i,j) \in \mathcal{X} \times \mathcal{U},
\end{gather}
where $\epsilon(t_a)$ is the error in the agent's estimate of the optimal Q-function.
\label{Thm: No Attack}
\end{thm}
We similarly derive the asymptotic limit of the
BDPMs when an attack is underway.

\begin{thm}
\label{Thm: Attack}
Given that the system is under attack, the BDPMs asymptotically limit as follows,
\begin{gather}
    \bar d_{i,j} \rightarrow  \hat Q(i,j) - Q^*(i,j) + \gamma  \mathbf{\hat p}_{i,j} \left(  \mathbf{v} - \mathbf{\hat v} \right)^T + \mathbf{\tilde p}_{i,j}(\mathbf{r}_{i,j} +  \gamma\mathbf{v})^T \\ \nonumber
    \text{a.s. as } t_c(i,j) \rightarrow \infty, 
    \; \forall (i,j) \in \mathcal{X} \times \mathcal{U}.
\end{gather}
\end{thm}

Similarly, we now prove a theorem that lower-bounds the magnitude of the BDPMs when the system is under attack.

\begin{thm}
\label{Thm: Attack Bound}
Given that the system is under attack, the magnitude of the BDPMs can be lower-bounded as follows,
\begin{gather}
    | \bar d_{i,j} | \geq \Phi_\text{min} - (1+\gamma) \epsilon(t_a),  \\ \nonumber 
  \forall (i,j) \in \mathcal{X} \times \mathcal{U}, \;\; \text{w.h.p. as } t_c(i,j) \rightarrow \infty.
\end{gather}
\end{thm}

Given that we have shown that the absolute value of the BDPM is upper-bounded when the system is secure and lower-bounded when the system is under attack, we can now derive conditions on when \algorithmref{alg:BellDev} will be able to detect MITM attacks. Intuitively, this happens when the safety upper bound is provably lower than the attack's lower bound, as formalized in the theorem below.

\begin{thm}
\label{thm:infoadv}
The information advantage condition,
\begin{equation}
\label{eq: info adv}
     \Phi_\text{min} > 2 \cdot (1+\gamma) \bar \epsilon,
\end{equation}
is necessary and sufficient for \algorithmref{alg:BellDev} to guarantee attack detection while avoiding false alarms with high probability as $t_c \rightarrow \infty$, when the agent uses any arbitrary policy. Here, $\bar \epsilon$ is an over-estimate of the agent's error in the Q-function, such that $\bar \epsilon \geq \epsilon(t_a)$.
\end{thm}
\subsection{Optimal System Identification}

While discussing model-free techniques to detect man-in-the-middle attacks on a system, one critical aspect to consider is the effectiveness of the agent's detection scheme, which is dependent on how accurately the adversary estimates the system model. In this section, we propose a system identification algorithm that the adversary can use and prove that it is optimal in terms of minimizing the deviation in the BDPM. 

Algorithm~\ref{alg:SysId}, which we refer to as the Sample Distribution Estimation algorithm, estimates the transition probability matrix by observing a training trajectory $\mathbf{\tau}_B$ to generate a frequency tensor $\mathbf{F}$. It then normalizes the entries of $\mathbf{F}$ to obtain the estimated transition probability matrix $\mathbf{\hat P}$. In Algorithm~\ref{alg:SysId}, $\mathbf{\tau}_B$ is a sequence of state-action pairs of length $t_b$, and $\mathbf{0}$ represents a zero tensor of appropriate dimensions. The algorithm explicitly iterates through each state $i \in \mathcal{X}$ and action $j \in \mathcal{U}$ to compute the transition probabilities to each subsequent state $k \in \mathcal{X}$. For state-action pairs $(i,j)$ observed in the trajectory, the transition probability is calculated as the empirical frequency of moving to state $k$ divided by the total number of times the pair $(i,j)$ was visited. For unvisited state-action pairs, a uniform prior is assigned to ensure $\mathbf{\hat P}$ remains a valid stochastic matrix.

\begin{algorithm}
\caption{Sample Distribution Estimation}
\label{alg:SysId}
\begin{algorithmic}
\REQUIRE $t_b \geq 0$, $|\mathbf{\tau}_B| = t_b$
\STATE \textbf{Initialize} $\mathbf{F} \in \mathbb{Z}^{|\mathcal{X}| \times |\mathcal{U}| \times |\mathcal{X}|} \gets \mathbf{0}$
\STATE \textbf{Initialize} $\mathbf{\hat P} \in \mathbb{R}^{|\mathcal{X}| \times |\mathcal{U}| \times |\mathcal{X}|} \gets \mathbf{0}$

\STATE // \textit{Count observed transitions}
\FOR{$n=1$ \TO $t_b-1$}
    \STATE $i \gets \mathbf{\tau}_B [n][0]$
    \STATE $j \gets \mathbf{\tau}_B [n][1]$
    \STATE $k \gets \mathbf{\tau}_B [n+1][0]$
    \STATE $\mathbf{F}[i,j,k] \gets \mathbf{F}[i,j,k] + 1$   
\ENDFOR

\STATE // \textit{Compute transition probabilities via explicit normalization}
\FORALL{$i \in \mathcal{X}$}
    \FORALL{$j \in \mathcal{U}$}
        \STATE $S \gets \sum_{k' \in \mathcal{X}} \mathbf{F}[i,j,k']$
        \IF{$S > 0$}
            \FORALL{$k \in \mathcal{X}$}
                \STATE $\mathbf{\hat P}[i,j,k] \gets \mathbf{F}[i,j,k] / S$
            \ENDFOR
        \ELSE
            \STATE // \textit{Handle unvisited pairs with uniform prior}
            \STATE $\mathbf{\hat P}[i,j,:] \gets 1/|\mathcal{X}|$ 
        \ENDIF
    \ENDFOR
\ENDFOR
\RETURN $\mathbf{\hat P}$
\end{algorithmic}
\end{algorithm}

To prove the optimality of the above simple statistical scheme, we must define the following key metric.

\begin{definition}[Asymptotic Bellman Deviation Process Gap]
    The asymptotic Bellman deviation process gap (BDPG) is the difference between the asymptotic limits of the BDPMs when the system is secure and when the system is under attack. Therefore,
    \begin{align}
        \lim_{t_c(i,j) \rightarrow \infty} \tilde d_{i,j} &= \lim_{t_c(i,j) \rightarrow \infty} \bar d_{i,j}^{\;\text{attack}} -  \bar d_{i,j}^{\;\text{secure}} 
        \\ \nonumber
        &= \mathbf{\tilde p}_{i,j} \left( \mathbf{r}_{i,j} + \gamma \mathbf{v} \right)^T 
    ,\forall (i,j) \in \mathcal{X} \times \mathcal{U}.
    \end{align}
\end{definition}

We can now show that \algorithmref{alg:SysId} is optimal in the sense of minimizing the Asymptotic BDPG given the adversary's learning trajectory $\boldsymbol{\tau}_b$. In other words, we show that using \algorithmref{alg:SysId} minimizes the BDPG up to fundamental statistical limits.

\begin{thm}
\label{MVUE}
    The sample distribution algorithm described in \algorithmref{alg:SysId} is an efficient estimator for the asymptotic BDPG. That is, $\mathbf{\hat P}$ satisfies the following,
    \begin{align}
        \mathbf{\hat p}_{i,j} &= \arg\min_{\mathbf{\hat{p}}_{i,j}|\boldsymbol{\tau}_b} \mathbb{E} \left[ |(\mathbf{p}_{i,j} - \mathbf{\hat p}_{i,j}) (\mathbf{r}_{i,j} + \gamma \mathbf{v})^T|^2\right], 
        \\ \nonumber
        &= \arg\min_{\mathbf{\hat{p}}_{i,j}|\boldsymbol{\tau}_b} \mathbb{E} \left[ |\tilde d_{i,j} |^2 \right]
        ,\forall (i,j) \in \mathcal{X} \times \mathcal{U}.
    \end{align}
\end{thm}

Finally, we prove that \algorithmref{alg:SysId} has the following asymptotic convergence bounds.

\begin{thm}
\label{thm:phi_conv}
    The sample distribution algorithm described in \algorithmref{alg:SysId} assures that,
    \begin{align}
    \label{eq:conv}
       |\mathbf{\tilde p}_{i,j} \left( \mathbf{r}_{i,j} + \gamma \mathbf{v} \right)^T| \xrightarrow[]{m.s} 0, \quad \forall (i,j) \in \mathcal{X} \times \mathcal{U},
    \end{align}
    as $t_b(i,j) \rightarrow \infty$, with a convergence rate of $O\left(\frac{1}{\sqrt{t_b(i,j)}}\right)$, where m.s refers to convergence in the mean-squared sense and $t_b(i,j)$ is the number of times the $(i,j)$ state-action pair occurred in the adversary's learning trajectory $\boldsymbol{\tau}_b$.
\end{thm}

\subsection{Extension to Dynamic Attacks}

The attack paradigm discussed in \sectionref{sec:PS} assumes synchronous attacks. In this section, we extend the paradigm to include asynchronous and dynamic attacks, while requiring little to no modification to the proposed Bellman deviation detection algorithm.

In the current paradigm, it is implicitly assumed that the adversary's attack phase begins simultaneously with the agent's detection phase. This ``synchronous" assumption is unrealistic in real-world scenarios. Therefore, we extend the paradigm to account for cases where the adversary initiates the attack before or after the agent begins detection. If the attack begins before detection, the setting is effectively identical to the synchronous case from the agent's perspective. Thus, we focus on the case where the attack begins after detection, with a finite lag $t_L$, and show that \algorithmref{alg:BellDev} maintains the same asymptotic guarantees.

\begin{thm}
    \label{thm:dyna_attack1}
    Given that the adversary's attack begins any finite lag $t_L$ after the start of detection. The Bellman deviation sequence has identical asymptotic limits as,
    \begin{equation}\nonumber
    \bar d_{i,j} \rightarrow  \hat Q(i,j) - Q^*(i,j) + \gamma  \mathbf{p}_{i,j} \left (  \mathbf{v} - \mathbf{\hat v} \right)^T, 
    \end{equation}
     when secure and
    \begin{equation} \nonumber
         \bar d_{i,j} \rightarrow  \hat Q(i,j) - Q^*(i,j) + \gamma  \mathbf{p}_{i,j} \left (  \mathbf{v} - \mathbf{\hat v} \right)^T + \mathbf{\tilde p}_{i,j}(\mathbf{r}_{i,j} +  \gamma\mathbf{v})^T
    \end{equation}
     when under attack, w.h.p as $t_c(i,j) \rightarrow\infty, 
    \; \forall (i,j) \in \mathcal{X} \times \mathcal{U}$.
    Similarly, the information advantage condition remains unchanged as
\begin{equation}\nonumber
    \Phi_\text{min} > 2\cdot (1+\gamma) \bar\epsilon.
\end{equation}
 Therefore, \algorithmref{alg:BellDev} guarantees attack detection while avoiding false alarms with high probability, as $t_c \rightarrow \infty $ even for asynchronous attacks, when the information advantage condition is met.
\end{thm}

We now extend the paradigm to include dynamic attacks, in which the adversary intermittently switches between attacking the system and allowing the agent to interact with the true environment. We show that the Bellman deviation detection algorithm can be slightly modified to guarantee attack detection w.h.p, and a low probability of false alarms.

To that end, we define the following key quantity for the adversary.
\begin{definition}[Adversary Attack Fraction]
    The attack fraction $\nu$ is defined as the asymptotic fraction of time the adversary spends attacking the system,
    \begin{equation}
        \nu = \lim_{t_c \rightarrow\infty}\frac{\sum_{t=1}^{t_c}I_a(t)}{t_c},
    \end{equation}
    where $I_a(t)$ is an indicator function that indicates if the adversary was attacking during time step $t$.
\end{definition}

We now show that the proposed algorithm can be modified slightly while still providing identical asymptotic guarantees. 

\begin{thm}
    \label{thm:dyna_attack2}
    Given that the adversary performs a dynamic attack with an attack fraction of $\nu$. The Bellman deviation sequence has the same asymptotic limits,
    \begin{equation}\nonumber
    \bar d_{i,j} \rightarrow  \hat Q(i,j) - Q^*(i,j) + \gamma  \mathbf{p}_{i,j} \left (  \mathbf{v} - \mathbf{\hat v} \right)^T, 
    \end{equation}
     when secure and a slightly different limit,
    \begin{equation} 
         \bar d_{i,j} \rightarrow  \hat Q(i,j) - Q^*(i,j) + \gamma  \mathbf{p}_{i,j} \left (  \mathbf{v} - \mathbf{\hat v} \right)^T + \nu \cdot \mathbf{\tilde p}_{i,j}(\mathbf{r}_{i,j} +  \gamma\mathbf{v})^T
    \end{equation}
     when under attack, w.h.p as $t_c(i,j) \rightarrow\infty, 
    \; \forall (i,j) \in \mathcal{X} \times \mathcal{U}$.
    Similarly, the information advantage condition is modified as,
\begin{equation}
    \nu \cdot \Phi_\text{min} > 2  (1+\gamma) \bar\epsilon .
\end{equation}
 Therefore, \algorithmref{alg:BellDev} guarantees attack detection  while avoiding false alarms with high probability as $t_c \rightarrow \infty $ even for dynamic attacks when the information advantage condition is met.
\end{thm}

\section{Cost-Tradeoff and Convergence Analysis}

So far, all analyses of attack detection guarantees have been limited to comparisons between the agent’s and the adversary’s error metrics. In this section, we formalize a method for analyzing detection efficiency in terms of the more tractable metric of learning times and provide guarantees on the detection efficiency of the proposed BDD algorithm. To that end, we first define the detection efficiency of an algorithm.

\begin{definition}(Detection Efficiency)\newline
Let $\mathcal{M} = (\mathcal{X}, \mathcal{U}, \mathbf{P}, \mathbf{R}, \gamma)$ be a Markov Decision Process.  
The detection efficiency \( \mathcal{E} \) of a reinforcement learning algorithm \( \mathcal{A} \) is defined as the minimum number of samples \( t_a(t_b, \delta) \) required for the agent to learn a policy \( \pi_{t_a} \) such that the probability of detecting the adversary who had learned for $t_b$ samples is \( 1 - \delta \)assuming both agent and adversary are sample efficient in learning. That is,
\[
t_a(t_b, \delta) = \min \{ t_a \mid \Pr ( \text{Attack Detection)} \geq 1 - \delta \}
\]

where \( \pi_{t_a} \) is the policy learned by the algorithm after \( t_a \) samples, the term $( \Phi_\text{min} > 2 \cdot (1+\gamma) \bar \epsilon)$ is the information advantage condition, $\Phi_\text{min}$ is the minimum value drift and $\epsilon$ is the agent's error in its Q-function.

Thus, a detection algorithm with lower \( t_a(t_b, \delta) \) is considered more detection efficient.
    
\end{definition}

There are fundamental information-theoretic limits on how efficient a detection algorithm can be. In the following lemma we derive an asymptotic universal lower bound on detection efficiency of any detection algorithm.

\begin{lemma}
    \label{lem:linear}
    The detection efficiency for any system $\mathcal{M}$ and any algorithm $\mathcal{A}$ is asymptotically lower-bounded linearly by $\Omega(t_b)$, as $t_b \rightarrow\infty$ and $\delta\rightarrow0$. 
\end{lemma}

Finally we now derive the asymptotic detection efficiency of the proposed BDD framework, and provide guarantees on its required learning time to maintain an information advantage on an adversary.
\begin{thm}
\label{thm:lineartime}
    The Bellman deviation detection algorithm (\algorithmref{alg:BellDev}) has an asymptotic detection efficiency bounded both below and above by $\Theta\left(\frac{t_b (1+\gamma)^2}{(1-\gamma)^3}\right)$, as $t_b \rightarrow\infty$ and $\delta\rightarrow0$.
\end{thm}

\begin{coro}
    \label{thm:ordereff}
    It follows from Lemma.~\ref{lem:linear} and \theoremref{thm:lineartime} that the proposed BDD algorithm is linear and order-optimal in its learning time, showing that it is the same order of efficiency in $t_b$ as model-based detection schemes.
\end{coro}

\section{Experimental Validation}
\subsection{Markov Decision Process Formulation}
We consider a finite Markov Decision Process (MDP) that models a stochastic random walk in a discrete one-dimensional state space. The MDP is represented as a tuple \( M = (S, A, P, R, \gamma) \), where:

\begin{itemize}
    \item \textbf{State Space} (\( S \)): The agent occupies a discrete state \( s \in S \), where \( S = \{-5, -4, \dots, 4, 5\} \). The boundary states \( s = -5 \) and \( s = 5 \) are absorbing states.  
    \item \textbf{Action Space} (\( A \)): The agent can take actions \( a \in A = \{-1, 0, 1\} \), corresponding to stepping left, staying in place, or stepping right, respectively.  
    \item \textbf{Transition Dynamics} (\( P \)): The transition probability function \( P(s' | s, a) \) defines the probability of transitioning to state \( s' \) given the current state \( s \) and action \( a \). The dynamics follow a stochastic motion model,
    
    \begin{equation}
    P(s' | s, a) =
    \begin{cases}
    0.05, & s' = s + a - 2 \\
    0.15, & s' = s + a - 1 \\
    0.60, & s' = s + a \\
    0.15, & s' = s + a + 1 \\
    0.05, & s' = s + a + 2 \\
    \end{cases}
    \end{equation}
    
    where the state transitions are absorbed at the boundaries, such that if \( s' \notin S \), the probability mass is assigned to the nearest valid state (i.e., \( -5 \) or \( 5 \)).  
    \item \textbf{Reward Function} (\( R \)): The agent receives a deterministic reward that decreases with the distance from the central state as, \begin{equation}
    R(s) =
    \begin{cases}
    5, & s = 0 \\
    4, & s \in \{-1, 1\} \\
    3, & s \in \{-2, 2\} \\
    2, & s \in \{-3, 3\} \\
    1, & s \in \{-4, 4\} \\
    0, & s \in \{-5, 5\}.
    \end{cases}
    \end{equation}
    
    \item \textbf{Discount Factor} (\( \gamma \)): The problem is formulated as an infinite-horizon discounted reward problem, where the agent maximizes the expected sum of discounted rewards,
        \begin{equation}
    J = \mathbb{E} \left[ \sum_{t=0}^{\infty} \gamma^t R(s_t) \right], \quad \gamma \in [0,1].
    \end{equation}
    For our experiments, we set the discount factor to $\gamma = 0.4$.
\end{itemize}

This MDP captures the noisy motion of a drunkard who attempts to move in a chosen direction but may overshoot or undershoot due to randomness. The structure of the reward function suggests that the agent should attempt to reach or remain near the central state \( s = 0 \), which provides the highest reward. 

\textbf{Agent Description:} The agent employs a traditional Q-learning algorithm as described in \cite{Q_learning}, with the learning rate $\alpha_n(i,j) = 1/n(i,j)$ and the $\epsilon$-greedy exploration strategy, where, $\alpha_n(i,j)$ is the learning rate for the $n$th step taken to learn $Q^*(i,j)$. The agent then uses the learned Q-function, $\hat Q(\cdot)$, to perform Bellman deviation detection as in \algorithmref{alg:BellDev}.

\textbf{Adversary Description:} The adversary learns the system optimally by estimating the sample distribution as described in \algorithmref{alg:SysId}. The adversary then uses the learned system estimate $\hat P$ to perform a MITM attack.

\subsection{Simulation}

We first validate the correctness of the algorithm by simulating single instances of the MITM attack problem. The agent is trained on 5 episodes (each 1000 samples long) and the adversary is trained on 1 episode of data, this raises the odds that the information advantage condition is met and that the agent will successfully detect an attack.

\begin{figure}[t]
\includegraphics[width=0.5\textwidth]{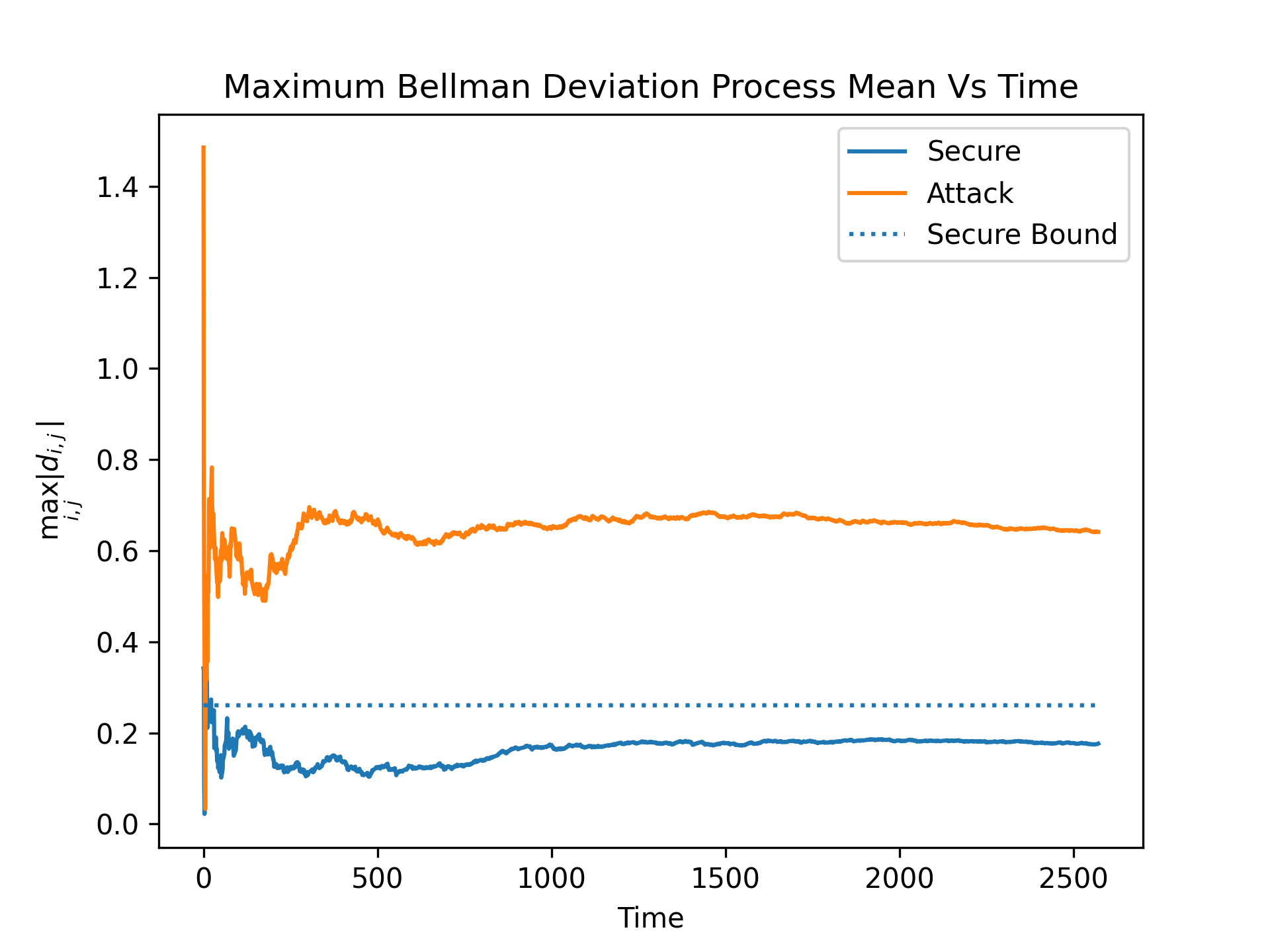}
\centering
\caption{This figure shows the absolute value of the maximum  BDPM for the cases when an attack occurs (in orange) and when the system is secure (in blue). The security bound (in dotted blue) upper bounds the BDPM when no attack occurs and lower bounds the BDPM when attack occurs, if the information advantage condition is met.}
\label{sing_inst}
\end{figure}

We now simulate two scenarios, one where the agent is not under attack and one where the adversary is performing a MITM attack. In both cases, the agent uses the proposed BDD algorithm to detect attacks by calculating the BDPM and comparing it to the security bound, given by $(1+\gamma)\epsilon.$ 
With high probability, the agent successfully detects an attack and the plots of the maximum BDPMs are compared in \figref{sing_inst}. As predicted by \theoremref{Thm: No Attack}, the absolute value of the maximum BDPM remains below the security bound when no attack occurs. Similarly, we observe that since the information advantage condition is met the absolute value of the maximum BDPM converges to a value over the security bound, as a consequence of \theoremref{Thm: Attack Bound} and \theoremref{thm:infoadv}. This validates the correctness of \algorithmref{alg:BellDev} in scenarios where the information advantage condition holds.

\begin{figure}[t]
\includegraphics[width=0.5\textwidth]{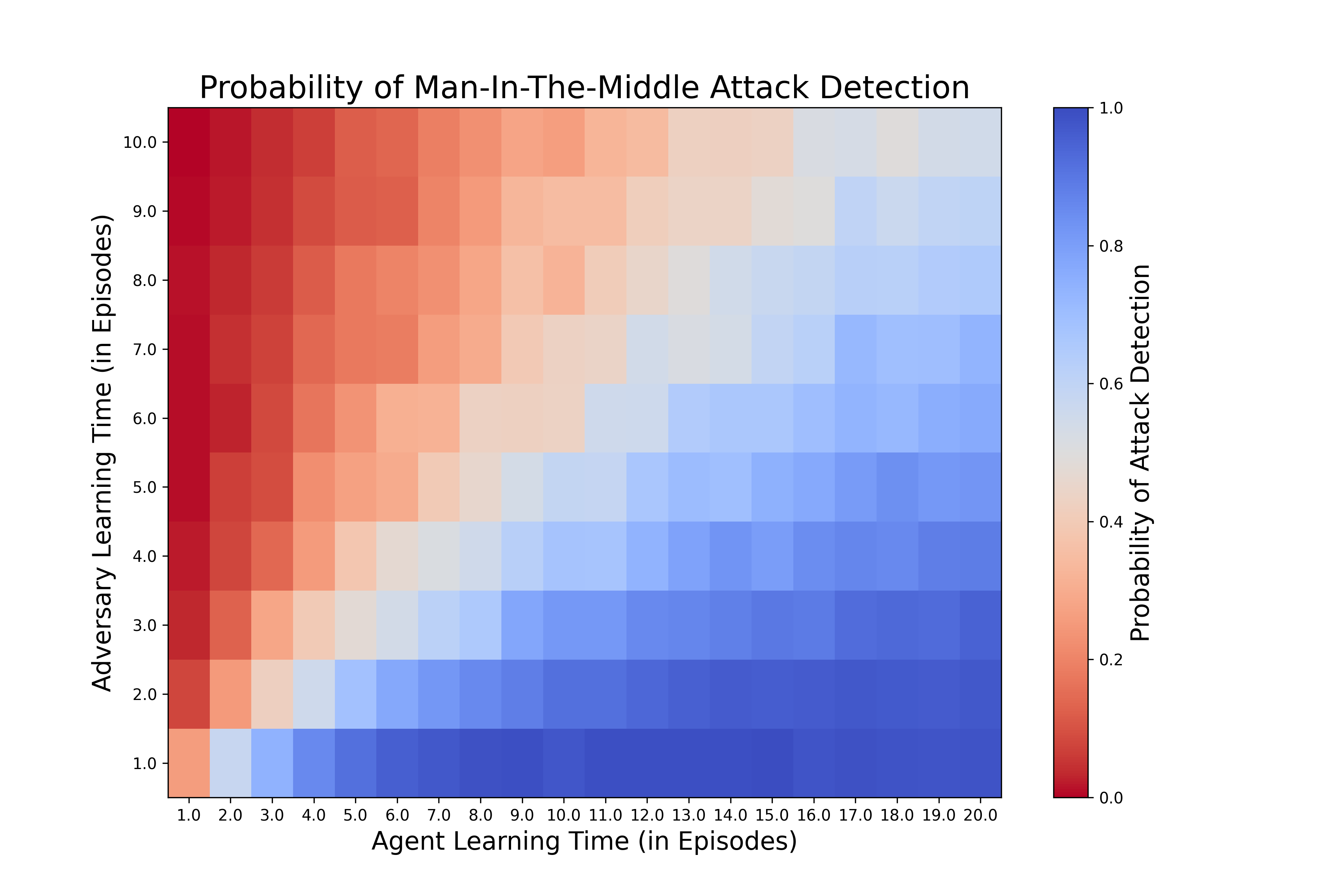}
\centering
\caption{This figure shows the probability of the agent detecting an attack, calculated based on 1000 trials for varying amounts of agent and adversary learning times. }
\label{fig:heatmap}
\end{figure}

To verify \theoremref{thm:lineartime}, we conduct Monte Carlo simulations to show that the minimum agent learning time required to detect an attack grows linearly with the adversary’s learning time, i.e.,$\Theta(t_b)$. To this end, we simulate the MITM attack problem for varying amount of agent and adversary learning times, measured in number of episodes. 1000 trials are conducted for each configuration of agent and adversary learning times. We then count the number of trials where the agent successfully detected the MITM attack underway and calculate the empirical probability of success. The results of the Monte-Carlo simulation are summarized in the heat map shown in \figref{fig:heatmap}.

As we expect the probability of attack detection rises as the agent learning times grows with respect to the adversary's, represented by the blue region. Similarly the probability of attack detection falls as the adversary learning time grows greater with respect to the agent's, represented by the red region. However the most notable feature of the heat map is that regions of high and low detection probability are separated by a distinct linear contour - a structure that is nontrivial and theoretically significant. While the contour could have been nonlinear, it is in fact linear, as predicted by \theoremref{thm:lineartime}. This result validates that the proposed BDD algorithm has linear asymptotic detection efficiency and hence order-efficient, as proven in \theoremref{thm:lineartime} and Corollary.~\ref{thm:ordereff}.

\section{Conclusion}
In this paper, we addressed the problem of securing reinforcement learning agents against Man-in-the-Middle (MITM) attacks in a model-free setting. By refining the Bellman Deviation Detection (BDD) framework to account for reward structures dependent on subsequent states, we captured the physical realities of control-loop interceptions where the adversary's estimation error manifests as system drift. We further demonstrated the robustness of our framework against dynamic attack strategies, including asynchronous and intermittent interceptions, proving that reliable detection is preserved even when the adversary attempts to remain hidden through sporadic activity. Our theoretical analysis established that the agent's learning time to secure the system is order-optimal, matching the adversary's own learning time $\Theta(t_b)$.

More broadly, the results show that meaningful security guarantees can be obtained even when the legitimate controller is fully model-free and has access only to learned value information rather than an explicit plant model. This is important because most existing detection guarantees in cyber-physical security are derived in model-based settings, where the defender is assumed to know or estimate the system dynamics directly. In contrast, our results show that attack detectability can still be characterized through Bellman-consistency deviations and that precise information-theoretic tradeoffs persist even in the absence of an explicit model. To our knowledge, this is among the first frameworks to provide rigorous attack-detection guarantees, dynamic-attack extensions, and asymptotic efficiency guarantees for learning-based MITM attacks in model-free reinforcement learning. As such, the paper helps bridge the gap between classical control-theoretic security and modern learning-based control, showing that security guarantees need not disappear when control is performed through reinforcement learning.

\section{Future Work}
While the present work establishes fundamental detectability limits and order-optimality for discrete MDPs, several promising directions remain. A primary objective is the extension of the BDD framework to high-dimensional and continuous state-spaces through the use of deep reinforcement learning and neural function approximators. Additionally, the application of this full framework to continuous-time linear systems, such as the Linear Quadratic Regulator (LQR) under model-free adaptive control, presents a significant open problem. Another important direction is the study of a dual safety problem, where post-training model drift or environmental change must be detected online using only model-free value information. Establishing analogous detectability guarantees in that setting would broaden the scope of the present framework beyond adversarial interception and toward real-time safety certification for learning-based control systems.

\section{Correctness of the Bellman Deviation Detection Algorithm}
\subsection{Proof of Theorem~\ref{Thm: lim1}}
\label{app:proof_theorem1}
We rearrange the terms of the Bellman equation \eqref{eq:BellmanEq} and subtract it from \eqref{eq:deviation average} to get:
\begin{align}
    \bar{d}_{i,j} = & \frac{\sum_{k=1}^{t_c(i,j)}  \hat{Q}(i,j) - r_{i,j}(x_{i,j}(k)) - w_{i,j}(k)} {t_c(i,j)}\\ \nonumber 
    & - \gamma  \frac{\sum_{k=1}^{t_c(i,j)}  
    \hat{V}(x_{i,j}(k))} {t_c(i,j)} - Q^*(i,j) + \mathbf{p}_{i,j} \left( \mathbf{r}_{i,j} +  \gamma \mathbf{v}\right)^T \\ \nonumber 
    = & \frac{\sum_{k=1}^{t_c(i,j)}  \left( \hat{Q}(i,j) - Q^*(i,j) \right) } {t_c(i,j)} -  \frac{\sum_{k=1}^{t_c(i,j)} w_{i,j}(k) } {t_c(i,j)} \\ \nonumber
    & -  \left( \frac{\sum_{k=1}^{t_c(i,j)}  r_{i,j}(x_{i,j}(k))} {t_c(i,j)} - \mathbf{p}_{i,j} \mathbf{r}_{i,j}^T \right) \\ \nonumber
    & - \gamma \left( \frac{\sum_{k=1}^{t_c(i,j)}  \hat{V}(x_{i,j}(k))} {t_c(i,j)} - \mathbf{p}_{i,j} \mathbf{v}^T \right).
\end{align}

The first term simplifies to $\hat{Q}(i,j) - Q^*(i,j)$. The second and third terms, due to the Law of Large Numbers (LLN), almost surely converge to zero. The fourth term converges to $\gamma  \mathbf{p}_{i,j} \left(  \mathbf{v} - \mathbf{\hat{v}} \right)^T$. Therefore, 
\begin{equation*}
    \bar{d}_{i,j} \rightarrow  \hat{Q}(i,j) - Q^*(i,j) + \gamma  \mathbf{p}_{i,j} \left(  \mathbf{v} - \mathbf{\hat{v}} \right)^T.
\end{equation*}

%\end{proof}
\subsection{Proof of Theorem~\ref{thm:UB}}
\label{app:proof_theorem2}
Given that $\bar{d}_{i,j}$ converges as:
$$ \bar{d}_{i,j} \rightarrow  \hat{Q}(i,j) - Q^*(i,j) + \gamma  \mathbf{p}_{i,j} \left(  \mathbf{v} - \mathbf{\hat{v}} \right)^T,$$
we use the convergence bound on the Q-function from \eqref{eq:ConvRate} to show that the first term involving $\hat{Q}(i,j) - Q^*(i,j)$ is bounded by $\epsilon_{t_a}$ from \eqref{eq:ConvRate}, and the second term, $\mathbf{p}_{i,j} \left(  \mathbf{v} - \mathbf{\hat{v}} \right)^T$, can be bounded as follows.

From \eqref{eq:ConvRate}, we know that 
\begin{gather}
 \label{eq: V bound}
  \left | \hat{V}(x) - V^*(x) \right | \leq \epsilon(t_a), \forall x \in \mathcal{X},
  \end{gather}
  and using this property, we get:
  \begin{gather}
   \gamma \mathbf{p}_{i,j} \left(  \mathbf{v} - \mathbf{\hat{v}} \right)^T \leq \gamma \sum_{\substack{x' \in \mathcal{X}}} p(i,j,x') \epsilon(t_a)  = \gamma \epsilon(t_a).
\end{gather}
Therefore, by using the triangle inequality, we can get that:
\begin{gather*}
    | \bar{d}_{i,j} | \leq (1+\gamma) \epsilon(t_a), \text{w.h.p. as } \\ \nonumber
     t_c(i,j) \rightarrow \infty, \;\; \forall (i,j) \in \mathcal{X} \times \mathcal{U}.
\end{gather*}
 \qedsymbol

\subsection{Proof of Theorem~\ref{Thm: Attack}}
\label{app:proof_theorem3}
In a manner similar to the proof of Theorem \ref{Thm: lim1}, we subtract equation \eqref{eq:BellmanEq} from \eqref{eq:deviation average} and introduce additional terms:
\begin{align}
    \bar d_{i,j} =& \frac{\sum_{k=1}^{t_c(i,j)}  \hat Q(i,j) - r_{i,j}(x_{i,j}(k)) - w_{i,j}(k)}{t_c(i,j)} \\ \nonumber
    & - \gamma \frac{\sum_{k=1}^{t_c(i,j)} \hat V(x_{i,j}(k))}{t_c(i,j)} - Q^*(i,j) + {r}_{i,j} \\ \nonumber
    & + \gamma  \mathbf{\hat p}_{i,j} \mathbf{v}^T + \gamma \mathbf{\tilde p}_{i,j} \mathbf{v}^T \\ \nonumber
    =& \frac{\sum_{k=1}^{t_c(i,j)}  \left( \hat Q(i,j) - Q^*(i,j) \right)}{t_c(i,j)} - \frac{\sum_{k=1}^{t_c(i,j)} w_{i,j}(k)}{t_c(i,j)} \\ \nonumber
    & -  \left( \frac{\sum_{k=1}^{t_c(i,j)}  r_{i,j}(x_{i,j}(k))}{t_c(i,j)} - \mathbf{p}_{i,j} \mathbf{r}_{i,j}^T \right) \\ \nonumber
    & - \gamma \left( \frac{\sum_{k=1}^{t_c(i,j)}  \hat V(x_{i,j}(k))}{t_c(i,j)} - \mathbf{\hat p}_{i,j} \mathbf{v}^T \right) + \gamma \mathbf{\tilde p}_{i,j} \mathbf{v}^T.
\end{align}

The first term simplifies to $\hat Q(i,j) - Q^*(i,j)$, the second term, due to the law of large numbers (LLN), almost surely converges to $0$, the third term converges to $\mathbf{\tilde p}_{i,j} \mathbf{r}_{i,j}^T$, and the fourth term converges to $\gamma \mathbf{\hat p}_{i,j} \left( \mathbf{v} - \mathbf{\hat v} \right)^T$. Therefore, 
$$ \bar d_{i,j} \rightarrow  \hat Q(i,j) - Q^*(i,j) + \gamma \mathbf{\hat p}_{i,j} \left( \mathbf{v} - \mathbf{\hat v} \right)^T + \mathbf{\tilde p}_{i,j} (\mathbf{r}_{i,j} + \gamma \mathbf{v})^T.$$
\qedsymbol
\subsection{Proof of Theorem~\ref{Thm: Attack Bound}}
\label{app:proof_theorem4}
Given that $\bar d_{i,j}$ converges as
$$ \bar d_{i,j} \rightarrow  \hat Q(i,j) - Q^*(i,j) + \gamma \mathbf{\hat p}_{i,j} \left( \mathbf{v} - \mathbf{\hat v} \right)^T + \mathbf{\tilde p}_{i,j} (\mathbf{r}_{i,j} + \gamma \mathbf{v})^T,$$
we use the convergence bound on the Q-function from \eqref{eq:ConvRate} to show that the first term, involving $\hat Q(i,j) - Q^*(i,j)$, is bounded by $\epsilon(t_a)$ from \eqref{eq:ConvRate}. The second term, $\mathbf{p}_{i,j} \left( \mathbf{v} - \mathbf{\hat v} \right)^T$, can be bounded as follows.

From \eqref{eq:ConvRate}, we know that
\begin{gather}
 \label{eq: V bound}
  \left| \hat V(x) - V^*(x) \right| \leq \epsilon(t_a), \forall x \in \mathcal{X},
\end{gather}
and using the above property, we get
\begin{gather}
   \gamma  \mathbf{\hat p}_{i,j} \left( \mathbf{v} - \mathbf{\hat v} \right)^T \leq \gamma \sum_{\substack{x' \in \mathcal{X}}} \hat p(i,j,x') \epsilon(t_a)  = \gamma \epsilon(t_a).
\end{gather}

Finally, the third term can be bounded using rigorous matrix algebra and norm inequalities:
\begin{equation}
 \left|  \mathbf{\tilde p}_{i,j} (\mathbf{r}_{i,j} + \gamma \mathbf{v})^T \right| \geq  \Phi_\text{min}.
\end{equation}

Therefore, using triangular inequalities, we prove that
\begin{gather*}
    | \bar d_{i,j} | \geq \Phi_\text{min} - (1+\gamma) \epsilon(t_a),  \\ \nonumber 
    \text{w.h.p. as } t_c(i,j) \rightarrow \infty. 
\end{gather*}
 \qedsymbol

\subsection{Proof of Theorem~\ref{thm:infoadv}}
\label{app:proof_theorem5}
From \theoremref{Thm: No Attack}, 
\begin{equation}
\label{eq:ub}
  | \bar d_{i,j}| \leq (1+\gamma ) \epsilon(t_a) \leq (1+\gamma) \bar\epsilon
\end{equation}
with high probability as $t_c \rightarrow \infty$, since $\bar\epsilon \geq \epsilon(t_a)$. Similarly, by \theoremref{Thm: Attack}, 
\begin{equation}
\label{eq:lb}
     |\bar d_{i,j}| \geq \Phi_\text{min} - (1+\gamma) \epsilon(t_a) \geq \Phi_\text{min} - (1+\gamma) \bar \epsilon 
\end{equation} 
with high probability as $t_c \rightarrow \infty$. Therefore, we can guarantee attack detection with no false alarms as $t_c \rightarrow \infty$ for \algorithmref{alg:BellDev}, if and only if
$$ \Phi_\text{min} - (1+\gamma) \bar \epsilon > (1+\gamma) \bar \epsilon.$$
That is, when the lower bound on the largest BDPM during an attack exceeds the upper bound on all BDPMs during no attack. This allows us to detect an attack when the lower bound is exceeded. We can now rewrite the above equation as
 $$ \Phi_\text{min} >  2 \cdot (1+\gamma) \bar \epsilon.$$
Since asymptotic attack detection with no false alarms with high probability can be achieved by \algorithmref{alg:BellDev} if and only if \eqref{eq: info adv} is true, this proves that \eqref{eq: info adv} is a necessary and sufficient condition.
\qedsymbol

\section{Optimal System Identification}
\subsection{Proof of Theorem~\ref{MVUE}}
\label{app:proof_theorem6}
Notice that minimizing $\mathbb{E} \left[ |(\mathbf{p}_{i,j} - \mathbf{\hat p}_{i,j}) (\mathbf{r}_{i,j} + \gamma \mathbf{v})^T|^2\right]$ is equivalent to finding the minimum variance unbiased estimator (MVUE) of $\mathbf{p}_{i,j}  (\mathbf{r}_{i,j} + \gamma \mathbf{v})^T$, which is a linear function of a row of $\mathbf{P}$.

Referring to Theorem 3 in~\cite{DSBai}, we know that any linear function on any row of $\mathbf{P}$ can be efficiently estimated by the same linear function over its sample distribution estimator $\mathbf{\hat P}$. Therefore, $\mathbf{\hat p}_{i,j}  (\mathbf{r}_{i,j} + \gamma \mathbf{v})^T$ is a MVUE of $\mathbf{p}_{i,j}  (\mathbf{r}_{i,j} + \gamma \mathbf{v})^T$. Hence, 
\begin{equation*}
    \mathbf{\hat p}_{i,j} = \arg\min_{\mathbf{\hat{p}}_{i,j}|\boldsymbol{\tau}_b} \mathbb{E} \left[ |\tilde d_{i,j} |^2 \right]
        ,\forall (i,j) \in \mathcal{X} \times \mathcal{U}. 
\end{equation*}
 \qedsymbol

\subsection{Proof of Theorem~\ref{thm:phi_conv}}
\label{app:proof_theorem7}
    In \theoremref{MVUE}, we proved that $\mathbf{\hat p}_{i,j}  (\mathbf{r}_{i,j} + \gamma \mathbf{v})^T$ estimated from \algorithmref{alg:SysId} is a MVUE, which implies that $\mathbb{E} \left[ \mathbf{\tilde p}_{i,j}  (\mathbf{r}_{i,j} + \gamma \mathbf{v})^T\right] = 0$, since
    \begin{align}
        \mathbb{E} \left[ (\mathbf{\hat p}_{i,j} - \mathbf{ p}_{i,j} ) (\mathbf{r}_{i,j} + \gamma \mathbf{v})^T \right] &= \mathbb{E} \left[ \mathbf{\tilde p}_{i,j}  (\mathbf{r}_{i,j} + \gamma \mathbf{v})^T \right] \\ \nonumber
        &= 0.
    \end{align}
    
    Similarly, the term $\mathbf{\tilde p}_{i,j}  (\mathbf{r}_{i,j} + \gamma \mathbf{v})^T$ can have its variance bound using the Cauchy-Schwarz inequality as
    \begin{equation}
    \mathbb{E} \left[ |\mathbf{\tilde p}_{i,j}  (\mathbf{r}_{i,j} + \gamma \mathbf{v})^T|^2 \right] \leq \mathbb{E} \left[ \|\mathbf{\tilde p}_{i,j}\|^2 \right] \cdot \|\mathbf{r}_{i,j} + \gamma \mathbf{v}\|^2.
    \end{equation}

    Since
    \[
    \mathbb{E} \left[ \|\mathbf{\tilde p}_{i,j}\|^2 \right] = \mathbb{E} \left[ \sum_{k=1}^N \tilde p_{i,j,k}^2 \right] \quad \text{and} \quad \mathbb{E} \left[ \tilde p_{i,j,k}^2 \right] = \frac{ p_{i,j,k} - p_{i,j,k}^2}{t_b(i,j)},
    \]
    which can be derived with elementary estimation theory.

    Hence,
    \begin{equation}
         \mathbb{E} \left[ |\mathbf{\tilde p}_{i,j}  (\mathbf{r}_{i,j} + \gamma \mathbf{v})^T|^2 \right] \rightarrow 0, \quad \text{as} \quad t_b(i,j) \rightarrow \infty,
    \end{equation}
    with a convergence rate of $O\left(\frac{1}{t_b(i,j)}\right)$. Since $\mathbf{\tilde p}_{i,j} \left( \mathbf{r}_{i,j} + \gamma \mathbf{v} \right)^T$ has a mean of 0 and its variance converges to 0, it converges to 0 in the mean-squared sense (m.s.).

    Now, we prove its convergence rate by invoking Chebyshev's inequality as
    \begin{equation}
    \label{eq:Cheb}
    |\mathbf{\tilde p}_{i,j}  (\mathbf{r}_{i,j} + \gamma \mathbf{v})^T|  \leq K \cdot \mathbb{E}\left[\|\mathbf{\tilde p}_{i,j} \|^2 \right]^{1/2} \cdot \|\mathbf{r}_{i,j} + \gamma \mathbf{v}\|, \quad \text{w.h.p.},
    \end{equation}
    for some large constant $K$. Since $\mathbb{E} \left[ \|\mathbf{\tilde p}_{i,j}\|^2 \right]$ converges to 0 as $O\left(\frac{1}{t_b(i,j)}\right)$, it follows that $\mathbb{E}\left[\|\mathbf{\tilde p}_{i,j} \|^2 \right]^{1/2}$ converges to 0 as $O\left(\frac{1}{\sqrt{t_b(i,j)}}\right)$.

    Using the above fact and \eqref{eq:Cheb}, this proves that \eqref{eq:conv} also has a convergence rate of $O\left(\frac{1}{\sqrt{t_b(i,j)}}\right)$.
\qedsymbol

\section{Extension to Dynamic Attacks}
\subsection{Proof of Theorem~\ref{thm:dyna_attack1}}
\label{app:proof_theorem8}

The fact that the asymptotic limit for the BDPM when no attack occurs is unchanged is trivially true. Since the start of the detection phase is of no consequence when no attack occurs. 
For the case of the asymptotic limit for the BDPM when an attack does occur, we decompose the BDPM series as follows,
\begin{align}
\bar{d}_{i,j} &= \frac{\sum_{k=1}^{t_c(i,j)} d_{i,j}(k)}{t_c(i,j)} ,\\ \nonumber
&= \frac{\sum_{k=1}^{t_d-1} d_{i,j}(k)}{t_c(i,j)} + \frac{\sum_{k=t_d}^{t_c(i,j)} d_{i,j}(k)}{t_c(i,j)} ,\\ \nonumber
&= \frac{t_d} {t_c(i,j)} \cdot \frac{\sum_{k=1}^{t_d-1} d_{i,j}(k)}{t_d}  + \frac{t_c(i,j) - t_d} {t_c(i,j)} \\ \nonumber 
&\cdot\frac{\sum_{k=t_d}^{t_c(i,j)} d_{i,j}(k)}{t_c(i,j) - t_d},
\end{align}
where $t_d$ is the time when the detection phase starts. As $t_c(i,j)\rightarrow\infty$, the two fractions converge as $\frac{t_d} {t_c(i,j)} \rightarrow 0$ and $ \frac{t_c(i,j) - t_d} {t_c(i,j)}\rightarrow 1$, thereby approaching the same limit as the asynchronous attack limit.
Finally since asymptotic limits for both cases remain the same \theoremref{thm:infoadv} similarly applies and we obtain the same information advantage condition.\qedsymbol

\subsection{Proof of Theorem~\ref{thm:dyna_attack2}}
\label{app:proof_theorem9}
The fact that the asymptotic limit for the BDPM when no attack occurs is unchanged is trivially true. Since the start of the detection phase is of no consequence when no attack occurs. 
For the case of the asymptotic limit for the BDPM when an attack does occur, we start by defining the attack and safe duration sets. Let $\mathcal{T}_a$ denote the set of time indices when an attack was occurring, i.e, $t\in \mathcal{T}_a \iff I_a(t)=1$. Similarly, let $\mathcal{T}_s$ be the complementary set, i.e, $t\in \mathcal{T}_s \iff I_a(t)=0$. We now, decompose the BDPM series, in the following manner, 
\begin{align}
\bar{d}_{i,j} &= \frac{\sum_{k=1}^{t_c(i,j)} d_{i,j}(k)}{t_c(i,j)} ,\\ \nonumber
&= \frac{\sum_{k\in \mathcal{T}_s} d_{i,j}(k)}{t_c(i,j)} + \frac{\sum_{k\in \mathcal{T}_a} d_{i,j}(k)}{t_c(i,j)} ,\\ \nonumber
&= \frac{|\mathcal{T}_s|} {t_c(i,j)} \cdot \frac{\sum_{k\in \mathcal{T}_s} d_{i,j}(k)}{t_d}  + \frac{|\mathcal{T}_a|} {t_c(i,j)}  
\cdot\frac{\sum_{k\in \mathcal{T}_a} d_{i,j}(k)}{t_c(i,j) - t_d}.
\end{align}
Now,  as $t_c(i,j)\rightarrow\infty$, the two fractions converge as $\frac{|\mathcal{T}_s|} {t_c(i,j)} \rightarrow 1-\nu$ and $ \frac{|\mathcal{T}_a|} {t_c(i,j)}\rightarrow \nu$. Hence the BDPM when under a dynamic attack has asymptotic limits as follows,
\begin{align} 
         \bar d_{i,j} &\rightarrow (1-\nu)(\hat Q(i,j) - Q^*(i,j) + \gamma  \mathbf{p}_{i,j} \left (  \mathbf{v} - \mathbf{\hat v} \right)^T ) \\\nonumber 
         &+ \nu (\hat Q(i,j) - Q^*(i,j) + \gamma  \mathbf{p}_{i,j} \left (  \mathbf{v} - \mathbf{\hat v} \right)^T 
         \\\nonumber 
         &+ \cdot \mathbf{\tilde p}_{i,j}(\mathbf{r}_{i,j} +  \gamma\mathbf{v})^T)
         \\ \nonumber 
         &=\hat Q(i,j) - Q^*(i,j) + \gamma  \mathbf{p}_{i,j} \left (  \mathbf{v} - \mathbf{\hat v} \right)^T 
         \\ \nonumber 
         &+ \nu \cdot \mathbf{\tilde p}_{i,j}(\mathbf{r}_{i,j} +  \gamma\mathbf{v})^T.
    \end{align}
Finally we apply \theoremref{thm:infoadv} with the new asymptotic limits to obtain the information advantage condition under a dynamic attack as,
\begin{equation*}
    \nu \cdot \Phi_\text{min} > 2  (1+\gamma) \bar\epsilon .
\end{equation*}
\qedsymbol

\section{Cost Tradeoff and Convergence Analysis}
\subsection{Proof of Lemma~\ref{lem:linear}}
By Theorem~\ref{thm:phi_conv}, after observing $t_b$ samples, the adversary can estimate each transition row with error of order
\begin{equation}
\left|\tilde p_{i,j}(r_{i,j}+\gamma v)^T\right| = O\!\left(\frac{1}{\sqrt{t_b(i,j)}}\right).
\end{equation}
Hence, the minimum detectable value drift generated by an optimally learning adversary can be as small as order $t_b^{-1/2}$.

On the other hand, by the information advantage condition, reliable detection requires
\begin{equation}
(1+\gamma)\epsilon(t_a) < \frac{\Phi_{\min}}{2}.
\end{equation}
Therefore, if the adversary has learned for $t_b$ samples and induces a drift of order $\Phi_{\min} = O(t_b^{-1/2})$, the agent must achieve value-estimation error
\begin{equation}
\epsilon(t_a) = O(t_b^{-1/2})
\end{equation}
in order to detect the attack with probability tending to one. Finally, for model-free reinforcement learning in discounted MDPs, achieving value error $\epsilon$ requires at least
\begin{equation}
t_a = \Omega(\epsilon^{-2})
\end{equation}
samples. Substituting $\epsilon = O(t_b^{-1/2})$ yields
\begin{equation}
t_a = \Omega(t_b).
\end{equation}
This proves the claim. \qedsymbol

\subsection{Proof of Theorem~\ref{thm:lineartime}}
\label{app:proof_theorem10}
Given that the adversary had $t_b$ to learn a policy, the goal is to find an upper bound on the minimum number of samples needed to achieve information advantage as, 
 \begin{gather}
        \Phi_\text{min} > (1+\gamma) \epsilon ,\nonumber \\ 
        \min_{i,j \in \mathcal{X}\times \mathcal{U}} |\mathbf{\tilde p}_{i,j} (\mathbf{r}_{i,j} + \gamma \mathbf{v}) |> (1+\gamma) \epsilon .
    \end{gather}
    However, in \theoremref{thm:phi_conv}, we proved that $ |\mathbf{\tilde p}_{i,j} \left( \mathbf{r}_{i,j} + \gamma \mathbf{v} \right)^T| \sim 
    {O\left(1/\sqrt{t_b(i,j)}\right)}$ and since $\min_{i,j} t_b(i,j) \sim O\left(\frac{t_b}{|\mathcal{X}||\mathcal{U}|}\right)$. We get, 
    \begin{equation}
        \label{eq:adv_learning_LB}
        O\left(\frac{\sqrt{|\mathcal{X}||\mathcal{U}|}}{\sqrt{t_b}}\right)> (1+\gamma) \epsilon .
    \end{equation}
    Therefore, $  \epsilon \sim O\left(\frac{\sqrt{|\mathcal{X}||\mathcal{U}|}}{\sqrt{t_b}(1+\gamma)}\right) $. However, we know that the lower bound on the sample complexity of Q-learning for the infinite-horizon discounted-reward problem is 
    $t_a \sim \Omega\left(\frac{|\mathcal{X}||\mathcal{U}|}{(1-\gamma)^3 \epsilon^2}\right),$ and that model-free RL algorithms do indeed achieve a sample complexity of 
    \begin{equation}
        \label{eq:samp_comp_LB}
        t_a \sim \Theta\left(\frac{|\mathcal{X}||\mathcal{U}|}{(1-\gamma)^3 \epsilon^2}\right),
    \end{equation}
    (refer to \cite{samp_comp_LB}). Therefore substituting 
    \eqref{eq:samp_comp_LB} in \eqref{eq:adv_learning_LB}, we finally get,
    \begin{equation*}
         t_a \sim \Theta\left(\frac{t_b (1+\gamma)^2}{(1-\gamma)^3}\right).
    \end{equation*}
 \qedsymbol

\bibliographystyle{IEEEtran}
\bibliography{ref.bib}
\end{document}